\documentclass{optica-article}

\journal{opticajournal} 

\articletype{Research Article}

\begin{document}

\title{Demonstration of an Integrated Planar Guided-wave Terahertz Synthesized Filter}

\author{Ali Dehghanian,\authormark{1,2} Mohsen Haghighat,\authormark{1,2}, Thomas Darcie\authormark{1}, and Levi Smith,\authormark{1,2,*} }

\address{\authormark{1}Department of Electrical and Computer Engineering, University of Victoria, Victoria, BC, V8P 5C2 Canada\\
\authormark{2}Centre for Advanced Materials and Related Technology (CAMTEC), University of Victoria, 3800 Finnerty Rd, Victoria, BC, V8P 5C2, Canada}

\email{\authormark{*}levismith@uvic.ca} 


\begin{abstract*} 
At terahertz (THz) frequencies there are few experimental works which demonstrate filter synthesis to obtain a desired filter response (i.e., Chebyshev, Butterworth, Bessel, etc.). Currently, the majority of literature perform THz filter analysis, that is, characterizing the filter response after design procedure. In this paper, we apply filter synthesis methods from microwave engineering to design several integrated planar low-pass filters \mbox{($f_c$ = 0.8 THz)}. We find that the transmission characteristics align with theory and simulation.
\end{abstract*}

\section{Introduction}
Filters are used throughout the electromagnetic (EM) spectrum for applications such as noise reduction, (de)multiplexing, and signal processing \cite{pozar_microwave_2011, oppenheim_signals_1997}. Filter design is a well established field for the electrical \cite{george_microwave_1980} and optical \cite{yariv_photonics_2006} frequencies. Within the terahertz (THz) gap (0.1-10 THz), filter design has historically focused on THz-bulk optical components which are used with radiated waves commonly associated with THz-time domain spectroscopy (TDS) \cite{valusis_roadmap_2021}. While these THz-bulk optical filters are useful, they do not integrate with miniaturized planar guided wave transmission lines (TLs) such as the coplanar striplines (CPS) and coplanar waveguides (CPW) \cite{Grischkowsky1988, cheng1994terahertz, Smith:21}. 

Microwave filter synthesis is applicable at THz frequencies; however, current literature is lacking validating experiments. This gap in literature likely originates from the difficulty associated with performing experiments at THz frequencies. To the authors knowledge, there are three main methods for characterizing a device-under-test (DUT) at frequencies above 1 THz while remaining inside the THz gap. First, electronic vector network analyzers (VNAs) are typically limited to less than 100 GHz, but this range can be increased using extender modules up-to 1.5 THz \cite{koller2016initial, Gao2021_THzfilt, cabello-sanchez_corrugated_2023, bauwens20141}. This method is attractive because it uses techniques that are commonplace in microwave engineering and does not require the use of photonics. However, the rectangular feedlines are dispersive and band-limited, thus many extension modules (costly) are required to perform broadband measurements. Next, THz optics (free-space radiation) can be used to to perform investigation of a DUT \cite{kim2010improvement, mbonye2009terahertz, atakaramians2013terahertz}. The source of THz radiation could be a photoconductive antenna or mixer, quantum cascade laser, free-electron laser, etc. \cite{lewis2014review}. The detector could be another photoconductive antenna, electro-optic crystal (ZnTe), Golay cell, bolometer, etc.\cite{lewis2019review}. THz optics provides valuable information about the DUT, but experiments are adversely impacted by field coupling errors which can be difficult to manage depending on the feedline geometry. Lastly, planar transmission lines (TLs) can be integrated with an transmitter and receiver used to investigate a DUT \cite{desplanque2004generation, kasai2009micro, potts2023chip, gomaa2020terahertz, smith_characterization_2021}. Typically these structures are integrated into a CPS or CPW TL with an shunt or series photoconductive switch. There are some key benefits using this method; namely, the structure is inherently compatible with planar structures, the propagating mode is quasi-TEM which does not exhibit significant dispersion, and the overall system cost is relatively inexpensive. In this work we investigate filters using photoconductive switches integrated with CPS TLs. Currently, this combination allows for characterization of DUTs up-to 3 THz \cite{smith_characterization_2021}.

In previous works, filter demonstrations at THz frequencies do not perform synthesis, that is, selecting the filter geometry to obtain a specific transfer function. Instead, authors perform filter analysis which generally involves designing a periodic device \cite{gomaa2020terahertz} or metastructure \cite{gil_metamaterial_2008, smith_characterization_2021, cabello-sanchez_capacitively-coupled_2023, guo_spoof_2018, xu_spoof_2020}, then confirming the transmission and reflection characteristics. While filter analysis is acceptable, it is generally advantageous to begin the design process by specifying targets such as cut-off frequency, minimal pass-band ripple, phase response, or roll-off rates. Here, we aim to demonstrate that the microwave filter synthesis methods are valid at THz frequencies. We note that it is not straightforward to characterize devices at THz frequencies due to limitations such as source and detector availability and signal loss from potential substrate radiation. We have had success using our THz system-on-chip (TSoC) platform that uses planar CPS TLs on a thin Si$_3$N$_4$ substrate to overcome these challenges \cite{smith_demonstration_2019}. This methodology enables device characterization from DC to frequencies beyond 3 THz and is compatible with several planar TL configurations. In this work, we integrate a synthesized Bessel stepped-impedance low-pass filters (LPF) into a CPS TL. The other TL configurations (i.e. CPW, slotline, etc.) are equally viable; however, the CPS configuration is best suited to work with photoconductive switches as a sliding contact source and detector \cite{Grischkowsky1988}.

The novelty of this work can be summarized as the first demonstration of an integrated planar guided-wave synthesised filter (Bessel) at THz frequencies. We note that we have selected a common filter topology (stepped-impedance LPF) for this proof-of-concept verification at THz frequencies.

\section{Design}
\begin{figure}[h]
    \centering
    \includegraphics[width=2.2in]{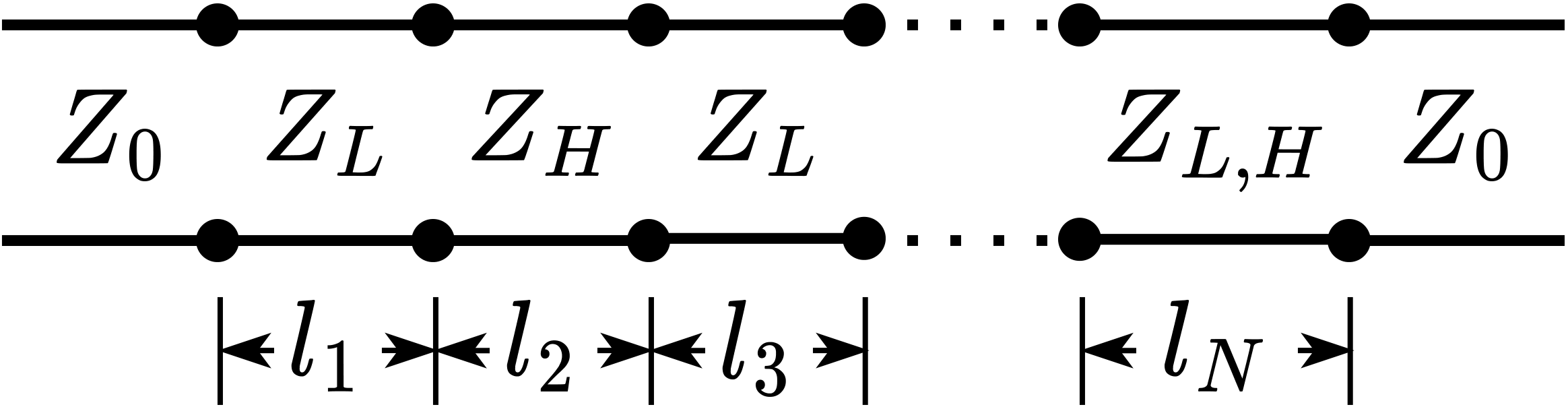}
    \caption{Stepped Impedance Filter}
    \label{fig:stepped_imp_filter}
\end{figure}

In this work we implement and test several linear-phase stepped-impedance low-pass filters on the TSoC. This filter consists of alternating sections of low and high characteristic impedance of varying lengths which depending on the specific filter configuration (Butterworth, Bessel, Chebyshev, etc.). Figure \ref{fig:stepped_imp_filter} illustrates this concept. Here, we focus on the Bessel filter because of the short physical length and minimal pulse distortion.

\begin{figure}[h]
    \centering
    \includegraphics[width=3.2in]{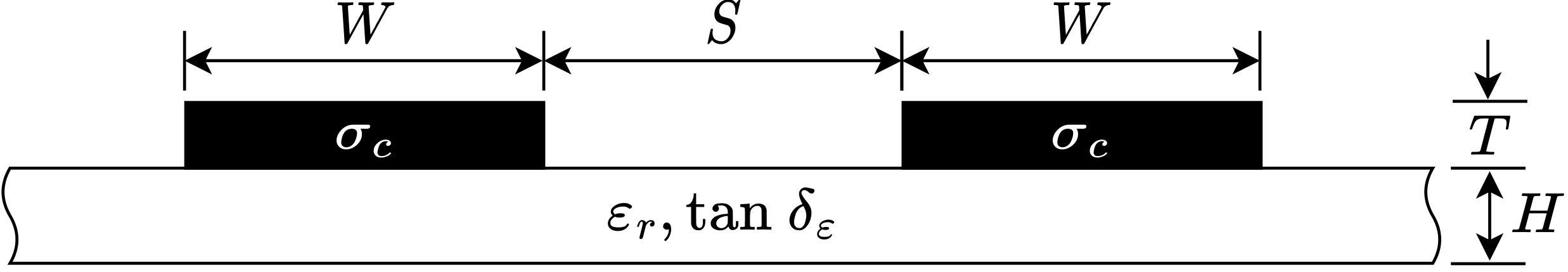}
    \caption{Coplanar strip transmission cross section}
    \label{fig:CPS_xsec}
\end{figure}

Figure \ref{fig:CPS_xsec} illustrates cross section of a CPS TL on a thin substrate. The conductor width, separation, thickness, and conductivity are given by $W$, $S$, \mbox{$T = 200$ nm}, and \mbox{$\sigma_c = 4.1\times 10^7$ S/m}, respectively. The selection of $W$ and $S$ is discussed below. The substrate thickness, relative permittivity, and loss tangent are given by $H=1\mu m$, $\varepsilon_r = 7.6$, and tan $\delta_e = 0.0056$, respectively \cite{cataldo_infrared_2012}. 

We are interested in the systematic design and experimental results of a low-pass Bessel filter which has a uniquely high cut-off frequency ($f_c > 500$ GHz). This can be demonstrated by fabrication on a thin substrate. The design procedure consists of specifying the desired cut-off frequency ($f_c$) and filter order ($N$), then selecting the appropriate feedline impedance ($Z_0$), low impedance ($Z_L$), and high impedance ($Z_H$), then calculating the section lengths ($l_n$).

\begin{figure*}[t]
    \centering
    \includegraphics[width=6in]{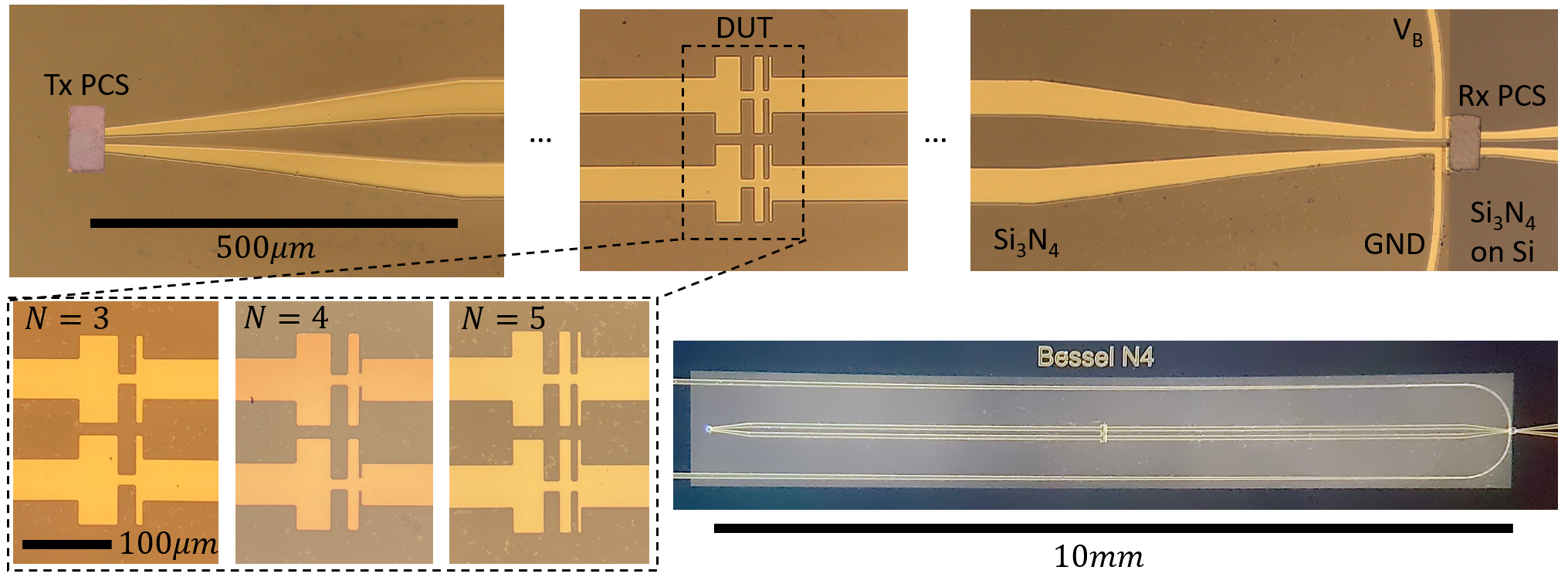}
    \caption{Microscope images of the TSoC structure with integrated filters.}
    \label{fig:fab_3rd_order}
\end{figure*}

In this work, we select $f_c = 0.8$ THz and fabricate 3rd, 4th, and 5th order filters ($N$ = 3, 4, 5). Figure \ref{fig:fab_3rd_order} illustrates the overall structure and the three different filters. The feedline impedance, $Z_0$, was selected to correspond to a low loss ($\alpha \approx$ 0.8 dB/mm \cite{Gomaa:20}) configuration which occurs when \mbox{$S_0 = 70 \mu m$} and \mbox{$W_0 = 45 \mu m$}. For a stepped impedance LPF, it is desired to maximize the $Z_H/Z_L$ ratio. Note that selection of $Z_L$ and $Z_H$ are subject to limitations. Also, we opted to keep the conductor center spacing constant ($S$+$W$=const.) throughout the filter length. Structures must have dimensions larger than 2 $\mu$m because of our photolithography resolution. At the upper end, given that we selected \mbox{$S_0 = 70 \mu m$} and \mbox{$W_0 = 45 \mu m$}, this implies $S+W=115\mu m$ and that $2\mu m<S_{L,H}<113\mu m$ and $2\mu m<W_{L,H}<113\mu m$.

In the past work \cite{smith_demonstration_2019}, we found that structures with $W=10 \mu m$ behave well with acceptable resistive loss. Thus, for this work, we selected our high impedance sections to have $W_H = 10 \mu m$ and $S_H = 105 \mu m$. For the low impedance sections, dielectric breakdown must be considered because the CPS TL also functions to DC bias the transmitting photoconductive switch. To ensure proper DC biasing, $S_L$ should be larger than $S$ at the transmitter ($S_{Tx} = 10 \mu m$). For this work, we have selected $S_L = 15 \mu m$ and $W_L = 100 \mu m$.

Next we must obtain $Z_0$, $Z_H$, and $Z_L$ which requires the use of full-wave simulations to obtain the impedance at the cut-off frequency ($f_c$ = 0.8 THz). Using ANSYS HFSS, we find $Z_0 = 234 \Omega$, $Z_L = 131 \Omega$, and $Z_H = 362 \Omega$.

Next we calculate the lengths of the individual sections. This is accomplished using well known methods \cite{pozar_microwave_2011}; we will quickly review the procedure here. Alternating low-impedance and high-impedance sections behave like alternating series inductors and shunt capacitors in a ladder circuit which exhibits low-pass behavior. The length of the individual sections are calculated using:
    \begin{equation}
        l_{L,n} = \frac{g_n}{\beta_c} \frac{Z_L}{Z_0}, \qquad l_{H,n} = \frac{g_n}{\beta_c} \frac{Z_0}{Z_H}.
        \label{eqn:sect_len} 
    \end{equation}
where $\beta_c \approx 2\pi f_c/c$. The filter element values, $g_n$, depend on the filter type (Bessel, Butterworth, Chebyshev, etc.) and order. For completeness, the element values are copied into Table \ref{tab:element_values} for the filters of interest \cite{pozar_microwave_2011}. The sections lengths for the different filter orders are calculated using (\ref{eqn:sect_len}) and tabulated into Table \ref{tab:sect_len}.

\begin{table}[h]
\renewcommand{\arraystretch}{1.3}
\caption{$g_n$ for maximally flat Time Delay LPF prototype \cite{pozar_microwave_2011}}
\label{tab:element_values}
\centering
\begin{tabular}{|c||c||c||c||c||c||c|}
\hline
N & $g_1$ & $g_2$& $g_3$& $g_4$& $g_5$ & $g_6$ \\
\hline
3 & 1.2550 & 0.5528 & 0.1922 & 1.0000 & - & - \\
4 & 1.0598 & 0.5116 & 0.3181 & 0.1104 & 1.0000 & - \\
5 & 0.9303 & 0.4577 & 0.3312 & 0.2090 & 0.0718 & 1.0000\\
\hline
\end{tabular}
\end{table}
\begin{table}[h]
\renewcommand{\arraystretch}{1.3}
\caption{Section lengths for filters}
\label{tab:sect_len}
\centering
\begin{tabular}{|c||c||c||c||c||c|}
\hline
N & $l_1 [\mu m]$  & $l_2 [\mu m]$& $l_3 [\mu m]$& $l_4 [\mu m]$& $l_5 [\mu m]$ \\
\hline
3 & 42 & 21 & 6  & - & - \\
4 & 35 & 20 & 11 & 4 & - \\
5 & 31 & 18 & 11 & 8 & 2  \\
\hline
\end{tabular}
\end{table}

\section{Experiment}
\begin{figure}[h]
    \centering
    \includegraphics[width=2.8in]{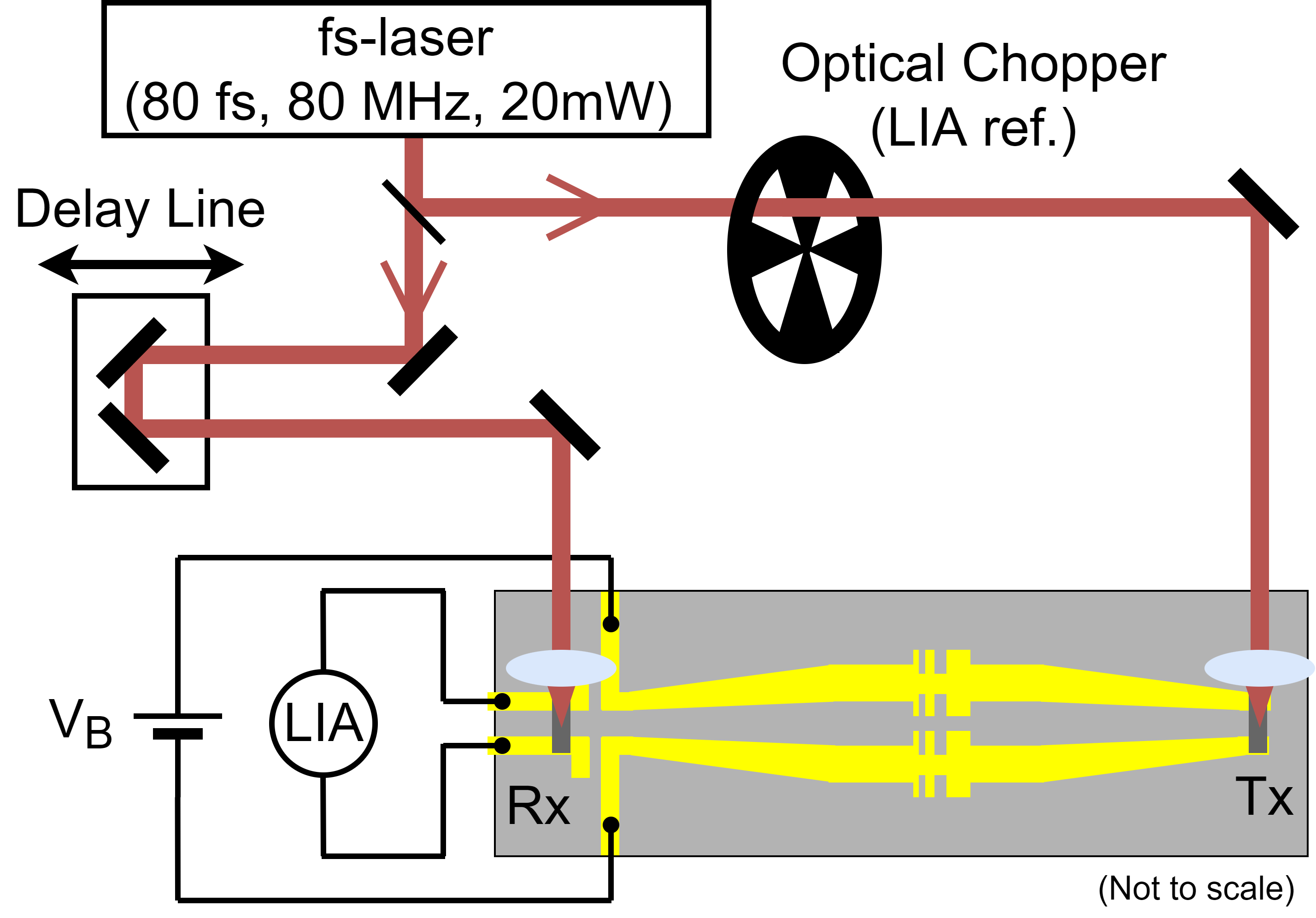}
    \caption{Simplified experimental setup.}
    \label{fig:experiment}
\end{figure}

To characterize the filters we use a method similar to our previous work \cite{smith_characterization_2021}. The experiment (Fig. \ref{fig:experiment}) consists of a modified THz-time domain spectroscopy (TDS) system which requires the use of electrical and optical components.

The transmitter is a thin (1.5 $\mu$m) DC-biased (24V) photoconductive switch (PCS) made of low-temperature grown gallium arsenide (LT-GaAs) that is Van der Waals (VDW) bonded to CPS TLs. A sub-picosecond (THz-bandwidth) electrical pulse is generated by illuminating the transmitter PCS with an 80 femtosecond (10 mW average) optical pulse from a mode-locked fiber laser. The electrical pulse propagates along the CPS TL and through the filters, or device-under-test (DUT), prior to arriving at the receiver. The receiver is another thin LT-GaAs PCS which is partially isolated from the transmitter circuit using a DC block. The receiver PCS is illuminated by the same optical pulse from the same mode-locked fiber laser, however, the pulse arrival time is made variable by the use of a mechanical delay line. Translation of the mechanical delay line reconstructs the transmitted signal which is given by the convolution of the incident electrical signal and receiver photoconductance which is measured by a lock-in amplifier (LIA) referenced to an optical chopper which modulates the optical beam in the transmitter path.

\section{Results and Discussion}
\begin{figure}[t]
    \centering
    \includegraphics[width=\textwidth]{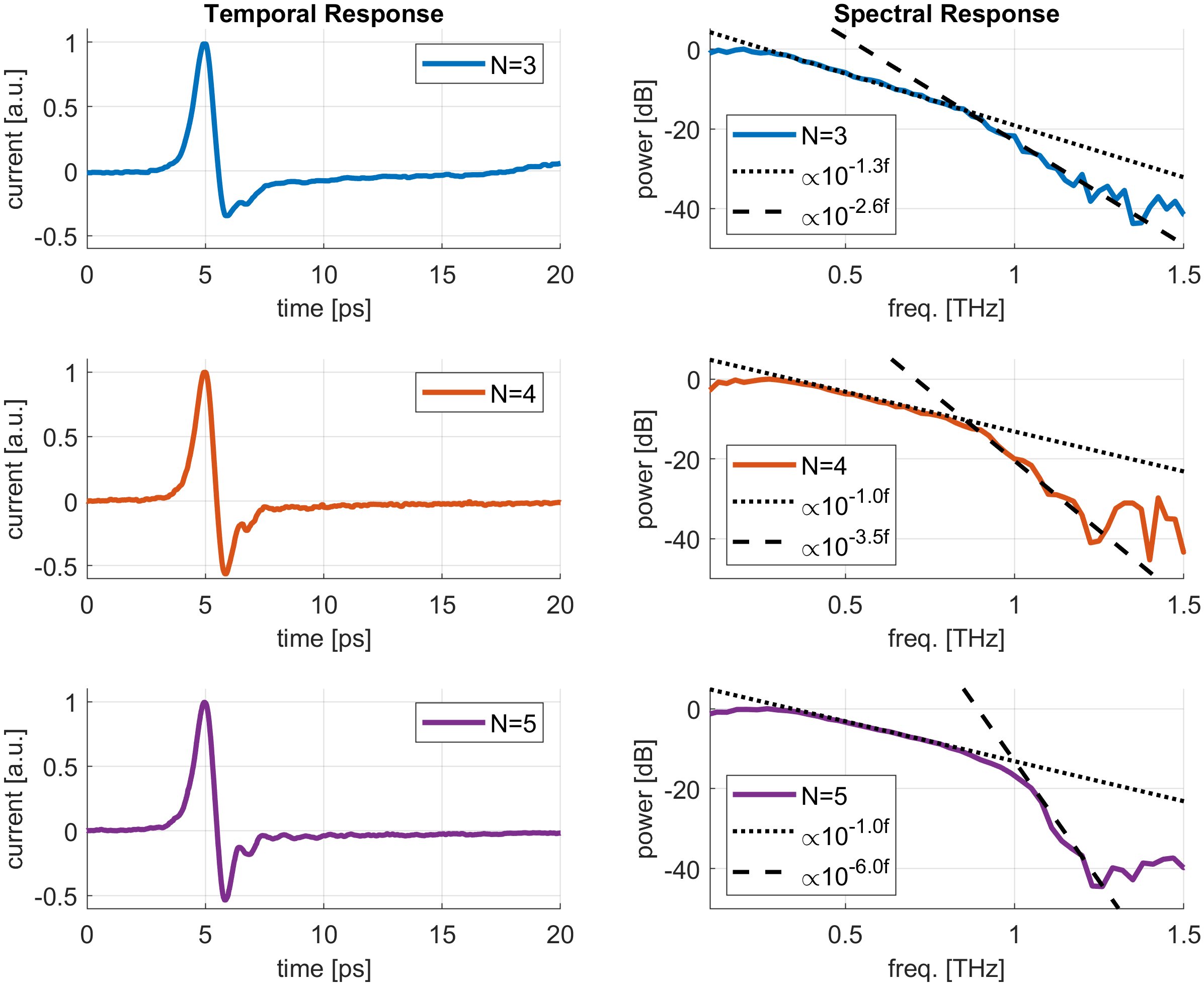}
    \caption{Experimental results for the N=3,4,5 Bessel filters.}
    \label{fig:bessel_results}
\end{figure}

Figure \ref{fig:bessel_results} plots the experimental temporal and spectral responses for the filters illustrated in Fig. \ref{fig:fab_3rd_order}. The temporal response illustrates minimal pulse distortion which is consistent with the linear phase response of the designed filters. This property is desired in a transient system because the ringing associated with a non-linear phase response can adversely impact the spectral resolution. The spectral response is obtained by applying the discrete Fourier transform to the temporal response. Recognizing that finite duration transient pulses do not have a flat spectral response, we expect an inherent roll-off. This concept is illustrated by the dotted lines in \mbox{Fig. \ref{fig:bessel_results}}. It is seen that the spectral response experiences a change of slope near the designed cut-off frequency at \mbox{0.8 THz}. This effect is illustrated by the dashed lines. The exponential coefficients for the dotted and dashed lines are noted in the legend of Fig. \ref{fig:bessel_results}. The increasing difference between the exponential coefficients for increasing filter order is expected since the higher order filters have a larger roll-off rate.

\section{Conclusion}
In this work we have designed, fabricated, and tested several stepped-impedance low-pass Bessel filters for THz applications. Characterization at THz frequencies was enabled by the use of our TSoC platform. We found that the experimental results align well with simulation and theory which illustrates the great potential for further experiment and developing synthesised low-pass filters. To the authors knowledge, this work demonstrates the first time a synthesized filter has been demonstrated in the THz gap.

\bibliography{main}

\end{document}